\documentclass[twocolumn,amsmath,aps]{revtex4}
\usepackage{epsfig, color, subfigure, amsmath, afterpage, url, float}
\usepackage[bookmarks]{hyperref}

\begin{document}

\title{The kSZ effect as a test of general radial inhomogeneity in LTB
  cosmology}

\author{Philip Bull}
\email{Phil.Bull@astro.ox.ac.uk}
\author{Timothy Clifton}
\email{Tim.Clifton@astro.ox.ac.uk}
\author{Pedro G. Ferreira.}
\email{P.Ferreira1@physics.ox.ac.uk}
\affiliation{Department of Astrophysics, University of Oxford, OX1 3RH, UK.}

\date{November 15, 2011}

\begin{abstract}
The apparent accelerating expansion of the Universe, determined from
observations of distant supernovae, and often taken to imply the existence of
dark energy, may alternatively be explained by the effects of a giant
underdense void if we relax the assumption of homogeneity on large
scales. Recent studies have made use of the spherically-symmetric,
radially-inhomogeneous Lema\^{i}tre-Tolman-Bondi (LTB) models to
derive strong constraints on this scenario, particularly from
observations of the kinematic Sunyaev-Zel'dovich (kSZ) effect which is
sensitive to large scale inhomogeneity. However, most of these previous studies
explicitly set the LTB `bang time' function to be constant, neglecting
an important freedom of the general solutions. Here we examine these models in
full generality by relaxing this assumption.  We find that although
the extra freedom allowed by varying the bang time is sufficient to
account for some observables individually, it is not enough to simultaneously
explain the supernovae observations, the small-angle CMB, the local
Hubble rate, and the kSZ effect.   This set of observables is strongly
constraining, and effectively rules out simple LTB models as an
explanation of dark energy.
\end{abstract}

\maketitle


\section{Introduction}\label{intro}

Observations of distant type-Ia supernovae are often taken to imply
that the Universe has entered a phase of accelerating
expansion, and may therefore contain `dark energy' \cite{perl,riess}. Such a
conclusion, however, cannot be inferred from the supernova data alone -- a 
model of the Universe is also required.
At present, the simplest and most widely applied cosmological models are
based on the Friedmann-Lema\^{i}tre-Robertson-Walker (FLRW) solutions of
general relativity. These solutions are highly
symmetric, and determining their validity as models of
the real Universe is of critical importance for determining the veracity of the claims involving dark energy. It is
toward these ends that this study is aimed: Are spatially homogeneous
and isotropic models with dark energy the only
ones capable of accounting for the recent cosmological observations that
appear to imply acceleration?

The inference of acceleration is of profound consequence, not just for
cosmology and gravitational physics, but also for particle and
high energy physics.  An accelerating Universe has an
entirely different causal structure from one that is decelerating, with the
vacuum itself taking on a non-zero energy density and becoming
thermal.  Beyond this there is the `cosmological constant
problem', that contributions from the zero-point energy of quantum
fields, and any bare cosmological term in Einstein's equations, must
cancel up to $1$ part in $10^{120}$  \cite{Weinberg}.  Such incredible
fine-tuning is widely believed to signify nothing less than a crisis
in modern physics, and so the task of verifying the assumptions that
go into our cosmological models becomes one of the utmost importance.

Here we focus on the problem of radial inhomogeneity, as modeled by
the Lema\^{i}tre-Tolman-Bondi (LTB) solutions of general relativity
\cite{LTB1,LTB2,LTB3}.  These are the general spherically-symmetric solutions of
Einstein's equations with dust.
They are widely known to have more than enough freedom to account
for the supernova observations without recourse to dark energy \cite{celerier},
and are often referred to as `void models' in the literature (but see
also \cite{Bolejko2}).  The relevant question is then whether or
not these models are compatible with other observational probes of
cosmology.

This question has been addressed by various authors
in a number of different contexts \cite{Biswas, Bolejko1, Bolejko3,
  CFL, GBH2, YNS, ZS, MZ11, MZS10, YNS2, Alnes, CFZ, ZMS}.  Most of these 
studies have, however, limited themselves to the special case of space-times that have a
spatially homogeneous energy density at early times.  This is
achieved by considering only those models that have a constant `bang
time'.  In this case it is known that while the small angle CMB
generated by a power-law spectrum of initial fluctuations can
be easily reproduced within void models, an unacceptably low value of $H_0$ is
required to do so \cite{ZMS,CFZ}.  However, it is also known that this problem can be alleviated by
allowing for {\it general} radial inhomogeneity, with non-constant
bang time \cite{CFZ}.  Here we address the problem of whether or not other
cosmological observables are also consistent with models that allow for this
additional freedom, as well as further investigating the parameter
space of solutions that fit the small-angle CMB.

We will be interested in particular in the kinematic
Sunyaev-Zel'dovich (kSZ) effect \cite{SZ80}. This effect is due to
the inverse Compton scattering of photons from the Cosmic Microwave
Background (CMB) off of electrons in distant clusters of galaxies.  The
rescattered light can be collected by observers who are
looking at the cluster. If in the rest frame of the electrons the CMB
has a non-zero dipole moment (in the direction of the observer) then the
reflected light that the observer sees has its spectrum shifted.  Such shifts are expected
to be observable by upcoming experiments, and although they have yet
to be directly detected \footnote{A detection of the kSZ effect is claimed in \cite{Kashlinsky1, Kashlinsky2}, but see also \cite{Osborne, Song, AtrioBarandela}.}, 
constraints have already been placed on the allowed magnitude of this effect \cite{kSZobs1,kSZobs2,kSZobs3,ACT,SPT}.

The kSZ effect is a particularly powerful probe of inhomogeneity as it
allows us to make observations not only along the null cone that is
the boundary of our causal past, but also along null curves that go inside this cone.  The
distant galaxy clusters essentially act as mirrors, reflecting light
from the last scattering surface that would otherwise be unobservable
to us.  This extra information is above and beyond that which is
available from the usual observations of distance measures, expansion
rates and number counts, and so it is of great potential significance
as a cosmological probe.  The power of the kSZ effect in this context
appears to have first been pointed out by Goodman in \cite{goodman},
although the first application of it to models that attempt to
account for dark energy was performed by Garc\'{i}a-Bellido and
Haugb{\o}lle \cite{GBH}.  These authors considered models with
constant bang time only.  We build on their work by allowing
for a radially dependent bang time.

To make progress it will be necessary to make a number of assumptions,
which to avoid confusion we will state here.  We assume the following:
\begin{itemize}
\item  That there is perfect spherical symmetry, with ourselves at the center of
  symmetry.
\item  That the formation of the last scattering surface proceeds
  as in FLRW cosmology.
\item  That there is a constant ratio of photons to baryons in the early
  universe.
\item  That the spectrum of initial fluctuations is a power law in wave number, $k$.
\item  That the energy density and all functions in the metric have smooth profiles.
 \label{caveats}
\end{itemize}
The first of these is inherent in the problem we have chosen to
address.  General perturbations to this exact set of symmetries have been
considered in \cite{CCF} and \cite{Zibin}, and the effects of being off-center have
been considered in \cite{Alnes}.  The second of these points is made
for convenience.  To date, we are unaware of any rigorous calculation
involving the formation of the last scattering surface in
inhomogeneous space-times.  The effect of allowing for an
inhomogeneous photon-to-baryon ratio has been considered in \cite{CR},
and the related question of changing the position of the last
scattering surface, while keeping the bang time constant, has been
addressed in \cite{YNS2}.  The effect of allowing a kink in the
spectrum of initial fluctuations has been considered in
\cite{Nadathur}.  We will not consider these freedoms further here,
but note that a constant bang time is an assumption that would
be added to similar lists in most other papers.  For details of the
effects of relaxing these assumptions, we refer the reader to the
papers cited above.

In Section \ref{model} we present the LTB solutions, and discuss how
distance measures and redshifts are calculated within them.  We then
discuss the effects of the two radial degrees of freedom in these
solutions, one of which is the bang time.  In Section
\ref{constraints}  we discuss some of the cosmological probes that can
be applied to these models, with particular reference to the kSZ
effect.  We also discuss why these observations are problematic for
LTB models with constant bang time.
In Section \ref{results} we present our results, which
include a detailed investigation of the effect of a radially varying
bang time on CMB and $H_0$ observations, as well as the kSZ effect.
We show that, despite the additional freedom in the bang time, there is a
combination of key observables that cannot be fitted simultaneously.
We conclude in Section \ref{discussion} that this effectively rules out
void models as an explanation of dark energy, unless one is prepared
to discard one or more of the assumptions that we have listed above.

\section{The model}
\label{model}

In order to model general radial inhomogeneity we will use the
Lema\^{i}tre-Tolman-Bondi (LTB) solutions of general relativity.
These are given by the line-element \cite{LTB1,LTB2,LTB3}
\begin{equation}
\label{LTBle}
ds^2 = dt^2 - {a^2_2(t, r) \over {1 - k(r)r^2}} dr^2 - a_1^2(t, r) r^2 d\Omega^2,
\end{equation}
where $a_2 = (a_1 r)^\prime$, and where $a_1$ must satisfy
\begin{equation}
\left ( \dot a_1 \over a_1 \right )^2 = {8 \pi G \over 3} {m(r) \over
  a_1^3} - {k(r) \over a_1^2}.
\label{LTBFriedmann}
\end{equation}
The functions $k(r)$ and $m(r)$ are arbitrary functions of the radial
coordinate, and primes and over-dots denote partial derivatives
with respect to $r$ and $t$, respectively. These solutions are exact,
and are the general spherically symmetric dust-only solutions to
Einstein's equations.  They admit a three dimensional group of
Killing vectors that act transitively on the surfaces of constant $r$
and $t$, and are spatially isotropic about the origin only.  

Solutions to Eq. (\ref{LTBFriedmann}) are of the form $a_1=a_1(r, t -
t_B(r))$, which introduces a third arbitrary function of $r$. This
gives a total of three free functions: $k(r)$, $m(r)$, and $t_B(r)$.
We refer to these quantities as the spatial curvature, gravitational mass 
density (distinct from the local energy density) and bang time, respectively. 
In the limit of homogeneity they are all constant.  It can also be seen that 
one can perform a coordinate transformation $r \rightarrow f(r)$ that
preserves the form of the metric in Eq. (\ref{LTBle}).  This freedom
can be used to set $m=$constant, without loss of generality (assuming 
$(mr^3)^\prime$ is always positive).  This leaves us with the general solution 
in terms of the spatial curvature, $k(r)$, and bang time function, $t_B(r)$, only. 
Analytic parametric solutions to Eq. (\ref{LTBFriedmann}) are known, and can be 
found in \cite{LTBsolutions}.

We can now use these solutions as cosmological models that exhibit
an arbitrary amount of radial inhomogeneity by supposing ourselves
to be observers at the center of symmetry.  Such models are known to be
able to produce excellent fits to the supernova data without requiring
any dark energy, and often result in the observer being at the center
of gigaparsec-scale underdensity, or `void'.  This is possible due
to both temporal {\it and} spatial variations in the geometry of
the space-time that are experienced by photons as they travel through the
void.  Such calculations require knowledge of redshifts and distance
measures in this space-time, which we will now consider.

Let us first define two different Hubble rates: a transverse one, $H_1
\equiv \dot{a}_1 / a_1$, and a radial one, $H_2 \equiv \dot{a}_2 / a_2$.  In the
limit of homogeneity these two quantities are identical, but differ, in
general, in inhomogeneous space-times.  The
redshift of photons traveling along radial geodesics can then be calculated by
integrating the radial Hubble rate as follows:
\begin{equation}
1+z = \exp \left\{ \int_{t_e}^{t_o} H_2(t,r(t)) dt \right\}, \label{eqn-redshift-central}
\end{equation}
where $r = r(t)$ is a solution of the radial geodesic equation, and
$t_e$ and $t_o$ are the time at which the photon was emitted and
observed, respectively.  Note that the relation $(1+z) \propto 1/a_1$ no
longer holds, in general.  We can also calculate the angular diameter
distance to objects at redshift $z$ using
\begin{equation}
d_A(z) = r(z) \; a_1(r(z), t(z)) , \label{eqn-da}
\end{equation}
where $t(z)$ is calculated by inverting $z=z(t)$ from
Eq. (\ref{eqn-redshift-central}), and $r(z)=r(t(z))$ is found using the radial
null geodesic equation.  Luminosity distances are then given by
Etherington's reciprocity theorem \cite{etherington}
\begin{equation}
d_L(z) = (1+z)^2 d_A(z),
\end{equation}
which is true in any space-time.  The local energy density is given by
$\rho = (m r^3)^\prime / 3 a_2 a_1^2 r^2$.

In this paper we will often choose to parametrize the two functions $k(r)$ and
$t_B(r)$ as Gaussian curves, with
\begin{eqnarray}
k(r) &=& A_k \exp(-{r^2 / \lambda^2_k}) + k_{\infty} \label{eqn-k-profile} \\
t_B(r) &=& A_{t_B} \exp(-{r^2 / \lambda^2_{t_B}}) f(r) , \label{eqn-tb-profile}
\end{eqnarray}
where $A_k$, $A_{t_B}$, $\lambda_k$, $\lambda_{t_B}$ and $k_{\infty}$ are
constants, and the factor $f(r)=\exp(-{r^{10} / \lambda^{10}_{t_B}})$ is
included to attenuate the bang time profile at large $r$.  This is
done so that wide profiles can be used, while limiting the effect of
early inhomogeneity on the central observer's last scattering surface, and is 
discussed further in Section \ref{tb}.  The profiles above are defined by their
amplitudes, $A_i$, and widths, $\lambda_i$.   A further parameter $k_{\infty}$ 
defines the asymptotic spatial curvature, outside the void.  The timescale of 
the model is set by choosing a local Hubble parameter 
$H_0=H_1\vert_{r=0}=H_2\vert_{r=0}$ at time $t_o$, where $(t_o-t_B)\vert_{r=0}$ 
is the age of the Universe along the worldline of an observer at $r=0$. A 
rescaling can be used to set $a_1(0, t_o) = a_2(0, t_o)= 1$.

The amplitudes in Eq. (\ref{eqn-k-profile}) and (\ref{eqn-tb-profile})
can be expressed in a more familiar form as fractions of the total
density at the origin today.  For this purpose let us define
\begin{eqnarray}
\Omega_{k_1} &\equiv& -A_k / H^2_0 \\
\Omega_{k_2} &\equiv& -k_{\infty} / H^2_0 \\
\Omega_k &\equiv& \Omega_{k_1}+ \Omega_{k_2}\\
\Omega_m &\equiv& {8 \pi G \over 3 H^2_0} m(r),
\end{eqnarray}
such that $\Omega_k + \Omega_m = 1$.
Furthermore, restricting ourselves to $\Omega_m \geq 0$ means that we consider only $\Omega_k \leq 1 $.

Let us now briefly consider the consequences of fluctuations in $k(r)$
and $t_B(r)$.  The former of these is the analog of the spatial curvature in FLRW
solutions, which dominates the dynamical evolution of the Universe at
late times.  The latter changes the location of the initial
singularity, and so can be thought of modifying the early stages of
the Universe's history.  This interpretation is supported by treating
the LTB geometry as a fluctuation about an FLRW solution.  In this
case the fluctuations in $k(r)$ can be mapped into growing modes,
while fluctuations in $t_B(r)$ are mapped onto decaying modes
\cite{Silk}.

\subsection{An Inhomogeneous Late Universe}

Let us first consider $k(r)$.  It can be seen from
Eq. (\ref{LTBFriedmann}) that $k<0$ gives a positive contribution to
the expansion of $a_1$, while $k>0$ gives a negative contribution.  This is the
behavior we are familiar with from the Friedmann equation of FLRW
cosmology.  Unlike the homogeneous FLRW solutions, however, the
expansion of $a_2$ does not always get a positive contribution from
$k<0$.  On the contrary, in regions of $k<0$ the expansion of $a_2$
can slow, and recollapse can occur.  This behavior is well known,
and can lead to the formation of a `shell crossing singularity' when
the collapsing region reaches $a_2=0$.

One can avoid shell crossing singularities in a region by satisfying the
Hellaby-Lake conditions \cite{HellabyLake}. These depend on the sign 
of $k(r)$, and for $k \le 0$ may be written using our notation as \cite{Sussman}
\begin{equation}
(m r^3)^\prime \ge 0, ~   t_B^\prime \le 0, ~~ {\rm and } ~~
(k r^2)^\prime \le 0,
\end{equation}
while for $k > 0$ the last of these should be replaced by
\begin{equation}
\left [ \log \left( {m \over k^{3 \over 2}}
    \right ) \right ]^\prime + {3 t_B^\prime |k| \over 8 \pi G m} \ge 0.
\end{equation}
These conditions guarantee that $a_2>0$, so that shell crossing
singularities cannot occur \footnote{There are some subtleties with
  the sign of $a_2$ when $k > 0$.  For details of this see
  \cite{Sussman}.}. They are, however, very restrictive, and most applications 
of the LTB solutions to cosmology simply avoid the issue by making sure that 
shell crossings only happen in the distant future.  They can then be considered 
as a breakdown of the model at some future time, after which a more 
sophisticated solution including pressure would be required to avoid the 
formation of singularities. The existence of pressure is expected to prevent 
the complete collapse of matter, and a large overdensity of collapsed 
structures is thus expected to form instead.

\subsection{An Inhomogeneous Big Bang}\label{tb}

Let us now consider the consequences of fluctuations in $t_B(r)$.  If $t_B(r)$
is not constant, then the `age of the universe' differs from place to
place.  This is a significant departure from the standard picture of the big bang, and
may initially seem odd. Certainly, there have been a number of objections to
allowing inhomogeneous bang times in the literature, with the result
that to date most studies of void models have expressly set
$t_B(r)=$constant {\it a fortiori}.  In this section we will discuss the
physical significance of an inhomogeneous big bang, and argue that it
is reasonable to consider models with such a feature.

As mentioned above, fluctuations in $t_B(r)$ correspond to decaying
modes when the space-time is approximated as a perturbed FLRW
solution.  A non-constant bang time therefore corresponds to an
inhomogeneous early universe, and as one goes further back in time the
size of the consequent inhomogeneity generally increases.  As with the
case of fluctuations in $k(r)$, there exist points beyond which the
scale factor $a_2$ is contracting rather than expanding, and in cases
where the Hellaby-Lake conditions are violated, shell crossing
singularities can occur.  In the case of fluctuations in $t_B(r)$,
however, this behavior occurs at very early times rather than at very
late times.  

\begin{figure}[htb]
  \centering
  \includegraphics[scale=0.65]{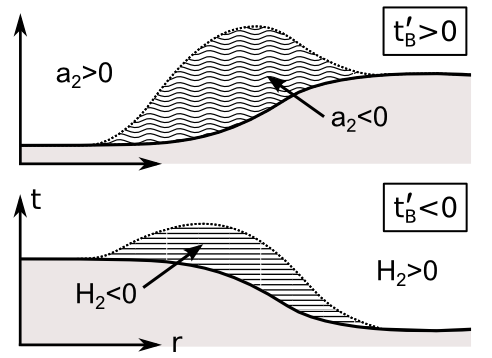}
  \caption{Upper panel: If $t_B^\prime > 0$ then a surface exists with
    $a_2=0$, corresponding to the occurrence of a shell crossing
    singularity (dotted
    line). Beyond this, the formal solution for the
    energy density gives negative values (hatched area). Lower
    panel: If $t_B^\prime < 0$, regions with $H_2 < 0$ form  (hatched
    area).  The solid lines correspond to the surfaces $t=t_B$.}
  \label{fig-tb-early-behaviour}
\end{figure}

In Fig. \ref{fig-tb-early-behaviour} we illustrate the existence of surfaces
of $a_2=0$ in cases with $t_B^\prime > 0$, and regions with $H_2 < 0$
in cases with $t_B^\prime < 0$.  The former of these are the early
universe analogue of the shell crossing singularities we described in
the previous section, and in this case we consider the singular
surface with $a_2=0$ to be our initial hypersurface.  The latter case
corresponds to regions that have started to collapse, but reach the
singular surface $t=t_B$ before shell crossings occur.  Contraction of
this type should be expected to cause blueshifts when looking along
exactly radial geodesics \cite{Hellaby}, and again this behavior has an
analog in the inhomogeneities that form at late times \cite{MZS10, Biswas}.
Such blueshifts would have profound effects if they were allowed to
occur between the last scattering surface and an observer (for example, 
a distance-redshift relation $r(z)$ that is not monotonic could occur).

C\'{e}l\'{e}rier {\it et al.} \cite{Bolejko2} found that small
variations in the bang time, of order hundreds of years, lead to
temperature anisotropies in the CMB of $\mathcal{O}(10^{-6})$,
which are currently only marginally too small to be observed.
Larger bang time variations would have a stronger observational
signal, but this need not be an issue if the region
of varying bang time occurs far inside the void, away from the surface
of last scattering that we see directly.  Nevertheless, observers
elsewhere in the space-time would see considerable anisotropies in
their CMB sky, and this could be observable in the kSZ effect we see
from CMB photons that rescatter off their cluster.  The kSZ effect
therefore has the potential to provide powerful constraints on the 
inhomogeneities caused by bang time fluctuations.

Of course, LTB models are dust-only solutions of Einstein's equations,
and lose their validity when radiation becomes important at early
times \cite{Bolejko2, CR}. The introduction of a radiation fluid into
inhomogeneous solutions complicates matters considerably, and we
do not attempt to include the gravitational effects of radiation in
our models here.  However, in the same way that one would expect pressure to prevent
the formation of shell crossing singularities at late times, one could
reasonably speculate on a similar mechanism occurring at early
times.  In the present study we concentrate on the matter dominated
phase of the Universe's history, which should be sufficient to model
the Universe from last scattering to the present time. This is well modeled by the
LTB solutions.  We leave the consideration of the gravitational effects of
inhomogeneous radiation fields to other studies.  For further details of this
the reader is referred to \cite{CR, Marra}.

An obvious concern with models of this type is that they are difficult
to reconcile with early universe inflation.  This is true with models
that have $t_B=$constant, as well as models with an inhomogeneous big
bang.  One must then either discard inflation for the time being, or
attempt to construct inflationary scenarios that result in occasional
large inhomogeneities (see, e.g., \cite{inf1} and \cite{inf2}).  Here we address
the problem of what can be said about the geometry of the Universe
directly from observations, rather than imposing requirements from
theories of the very early universe.

\section{Observational probes, and results with constant bang time}\label{constraints}

Void models that reproduce the observed supernova distance modulus
curve have proved relatively easy to construct, and little more than a
moderately deep underdensity with a comoving width of order a
gigaparsec is required to obtain a satisfactory fit.  Indeed, the ease
with which the supernova observations can be reproduced has been one
of the principle factors motivating interest in these models.

The introduction of such a large inhomogeneity, however, can hardly be
expected to leave predictions for other cosmological observables
unchanged, and so there have been a number of attempts to make
detailed tests of voids using multiple data sets \cite{Biswas, ZMS,
  GBH2}.  While thorough, these previous studies have limited
themselves to the case of voids with constant bang times.
In this section we summarize the constraints that can be imposed on
such void models from observations of supernovae, the CMB, the local
Hubble parameter, and the kSZ effect.  In particular, we draw
attention to the difficulty that voids with constant bang time have in
fitting the CMB and $H_0$ simultaneously, and discuss the
power of the kSZ effect as a test of large-scale inhomogeneity in
these models \cite{GBH, ZS, MZ11}.

In Section \ref{results} we will proceed to consider more general
voids with varying bang times.  This additional freedom allows some of
the constraints on the specific observables discussed in this section
to be weakened significantly (although a combined fit to all data sets
remains elusive).

\subsection{Supernovae}\label{constraints-sne}

As noted above, fitting the supernova data is a relatively simple matter,
and void models can be constructed that fit any given $d_L(z)$ curve
\textit{exactly} \cite{MHE}.  One should be aware, however, that reproducing 
the precise effects of $\Lambda$ at low $z$ requires an
energy density profile that is `cusped' at the center \cite{CFL,FLSC}.
Generic smooth profiles produce qualitatively different behavior,
due to the Milne-like geometry near the origin, but can still be shown
to be consistent with current data sets \cite{CFL,FLSC}.  

\subsection{The CMB and $H_0$}\label{constraints-cmb}

If the last scattering surface we observe is located in a region
of the Universe that is homogeneous and isotropic enough to be
modeled as being approximately FLRW then we can use standard
techniques to calculate the power spectrum of fluctuations on that
surface.  The CMB that we measure on our sky then depends on the
initial spectrum, which can be calculated using an effective FLRW
model, and the projection of fluctuations from the last scattering surface onto our
sky.   This projection depends on the space-time geometry between us and the
last scattering surface, and can be calculated from the angular diameter
distance in Eq. (\ref{eqn-da}).  It is in this way that the CMB
provides constraints on the geometry of the late Universe.

In general, voids and FLRW models with the same local geometry have
different angular diameter distances to the last scattering surface,
resulting in a relative shift in their observed CMB power spectra.
Now, the distance to last scattering can be adjusted by changing the
width and depth of the void, but this typically produces relatively
small shifts that are not enough to bring the peak positions of the
CMB power spectrum in line with current observational constraints
\cite{CFZ}. Changing the curvature of the FLRW region near the last
scattering surface, however, produces much larger effects \cite{CFZ}, and good fits to the
small-scale CMB power spectrum can be found for void models
that have positive asymptotic curvature.  Such models, however,
require an anomalously low local Hubble rate ($H_0 \lesssim 50$ km
s$^{-1}$ Mpc$^{-1}$) in order to keep the expansion rate at last
scattering low enough to be consistent with the data \cite{ZMS,CFZ}.
This is strongly inconsistent with the observed value of $H_0 = 73.8 \pm 2.4$
km s$^{-1}$ Mpc$^{-1}$ recently found in \cite{Riess}.
The CMB$+H_0$ by themselves are therefore sufficient to effectively
rule out simple void models with constant bang time.

One can attempt to avoid this conclusion by violating the assumptions
that we set out in Section \ref{intro}.  In particular, models with 
inhomogeneous last scattering surfaces have been
considered in \cite{YNS} and \cite{CR}, and a non-power law spectrum of
initial fluctuations has been considered in \cite{Nadathur}.  If one
is prepared to consider such additional freedoms then the CMB$+H_0$
constraints can be considerably weakened.

In Section \ref{results} we will consider the consequences of CMB
observations in general void models, where the bang time is allowed to vary.
It has already been shown in \cite{CFZ} that the available constraints
from the CMB$+H_0$ can be considerably weakened in this case,
without any need to violate the assumptions introduced in Section \ref{intro}.
In Section \ref{results} we will quantify this result, finding best
fit models and confidence regions in parameter space.  We will
show that the best fit models have bang time fluctuations of order a
billion years.

\subsection{The kSZ Effect}\label{ksztest}

A promising observable for testing large-scale homogeneity is the kinematic
Sunyaev-Zel'dovich (kSZ) effect.  This effect occurs because galaxy clusters contain hot gas that
can Compton scatter CMB photons, leading to a frequency-dependent
temperature increment/decrement of the CMB along the line of sight to
the cluster.  Such scattering events cause two separate effects that
can be distinguished in the reflected light.  The first is due to transfer of thermal energy from
the cluster gas to the photons, and is known as the thermal Sunyaev-Zel'dovich
effect.  The second is due to the dipole, $\Delta T/T$, of the CMB radiation on the
sky of an observer comoving with the reflecting cluster, and is known
as the kinematic Sunyaev-Zel'dovich effect \cite{SZ80}.  It produces a temperature
change of $\Delta T / T$ in the reflected light, and is similar to
the relativistic Doppler shift that one would experience by reflecting
a beam of light off of a moving mirror.

As noted by Goodman \cite{goodman}, and explicitly calculated by 
Garc\'{i}a-Bellido and Haugb{\o}lle \cite{GBH}, observers who are off-center 
in a radially-inhomogeneous universe should expect to see a large
$\Delta T/T$ in their CMB if they are comoving with the dust. This is because the 
distance-redshift relation becomes a function of direction in an inhomogeneous 
space-time, so that the surface of last scattering will appear to be at different 
distances/redshifts in different directions on the sky of an
off-center observer.  The result of a large kSZ effect then follows
because most observers in the space-time see a highly anisotropic CMB.
Of course, this is not the case in an FLRW universe, where one should
anticipate a low kSZ signal due only to the peculiar motion of clusters.
It is for this reason that the kSZ effect is expected to be a powerful
probe of large-scale inhomogeneity.

In a void model the dipole in the CMB is aligned in the radial
direction due to spherical symmetry, and can be calculated from the
relative velocity to an observer at the same point who would see an
isotropic CMB.  It is this dipole that can then, in principle, be 
measured using the kSZ effect. Now, the magnitude of dipole, $\Delta T/T$, can be
calculated for a given void model by finding the redshifts to last
scattering when looking radially into and out of the void.  The
observer then sees an average temperature of $T={1 \over 2}(T_{in} +
T_{out})$, where $T_{in}$ and $T_{out}$ are the temperatures of CMB
photons seen when looking into and out of the center of symmetry,
respectively, and the relative velocity with respect to the CMB rest
frame that causes this dipole is given by
\begin{equation}
\beta = {\Delta T \over T}  = {{z_{in} - z_{out}} \over {2 + z_{in} + z_{out}}},
\label{ksz-dt}\end{equation}
where $z_{in}$ and $z_{out}$ are the redshifts to last scattering in
the directions toward and away from the center of the void,
respectively.  They can be calculated from
Eq. (\ref{eqn-redshift-central}), which is valid for off-center
observers \cite{CZuntz}.
The kSZ effect can also be measured as a power spectrum using \cite{ZS, MZ11}
\begin{equation}
\left . {\Delta T (\hat{n}) \over T} \right |_{kSZ} = \int_0^{z_*}
\beta(z) \delta_e(\hat{n}, z) {d\tau\over dz} dz,
\label{eqn-ksz-power} 
\end{equation}
where $\Delta T(\hat{n}) / T$ is the CMB anisotropy seen by an
off-center observer in a direction $\hat{n}$, $\tau$ is the optical depth along
the line of sight, $\delta_e$ is the density contrast of
electrons, and $z_*$ is the redshift to the last scattering surface.

One should note that the calculation described above over-estimates
$\Delta T/T$ in the reflected light because the anisotropy seen by off-center observers will not
be purely dipolar, especially far
from the center of the void \cite{Alnes}.  The dipole
contribution, however, is the dominant one at low $z$, and so we
expect the prescription outlined above to be accurate enough for our
current purposes.  Observations of individual clusters have yielded upper 
limits of $\Delta T/T \lesssim 2000$ kms$^{-1}$ \cite{GBH}, and more recent observations 
from ACT and SPT have produced upper limits on the kSZ power spectrum 
at $\ell = 3000$ of 8 $\mu$K$^2$ and 13 $\mu$K$^2$, respectively \cite{ACT, SPT}. 
This is consistent with the typical peculiar
velocities expected in $\Lambda$CDM of $\sim$400 kms$^{-1}$, but is
strongly inconsistent with any large void with constant bang time that
obeys the assumptions made in Section \ref{intro} \cite{GBH}. 

\begin{figure}[htb]
  \centering
  \includegraphics[scale=0.45]{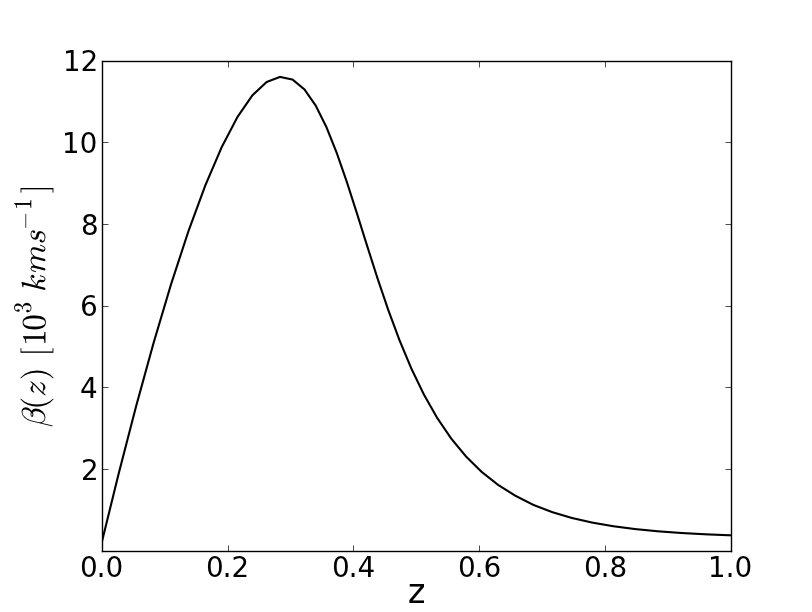}
  \caption{An example of the relative velocity with respect to the CMB
    rest frame, $\beta(z)$, that
    would be inferred by a central observer in an example giant void
    with constant bang time.}
  \label{fig-vp-basic}
\end{figure}

An example $\beta(z)$ profile that an observer at the center of a
large void with constant bang time could infer from observations of
the kSZ effect is shown in Figure \ref{fig-vp-basic}.  Although this
is only one example, the enormous magnitude of the effect is a
generic result for observers located at the center of such voids.
This directly demonstrates the utility of the kSZ effect as a probe of
inhomogeneity on large scales, and explains why current observational constraints on
the kSZ effect by themselves are enough to rule out simple voids with constant bang
time.  In Section \ref{results} we consider the consequences of a
varying bang time on observations of the kSZ effect, and show that
there exist general giant void models that are consistent with
constraints from current observations.

\subsection{Other Observables}
\label{obs-others}

So far we have discussed the specific observables of
supernovae, the CMB$+H_0$, and the kSZ effect, and described how the latter two
of these provide constraints on simple void models with a constant bang time
that are sufficient to effectively rule them out.  We have chosen to
discuss these particular observables as we consider them to be
reasonably well defined in void models, easily calculable, and very
constraining.  There are, of course, other observables that one could
also consider.  These include baryon acoustic oscillations (BAOs),
galaxy correlation functions, and
the integrated Sachs-Wolfe (ISW) effect, to name just a few.

To understand these observables in the FLRW cosmological
models one only needs linear perturbation theory about an FLRW
background, which is very well understood.
The observables in question can therefore be straightforwardly calculated.  Linear
perturbation theory in LTB cosmology, however, is significantly more
complicated.  In \cite{CCF} a gauge invariant formalism for
general perturbations in spherically symmetric space-times is applied to
these models, and it is shown that scalar, vector and tensor modes no
longer decouple.  This means that complicated effects can occur that
are not present in FLRW cosmology.  The full consequences of this
behavior have yet to be understood, and so here we avoid the use of
observables that rely on linear perturbation theory.  This includes
BAOs, galaxy correlation functions and the ISW effect.  For treatments
of some of these observables when $t_B=$constant the reader is
referred to \cite{MZS10}.

\section{Results with varying bang time}
\label{results}

In this section we examine the constraints that can be imposed on
general void models in which the bang time is allowed to vary.  This
generalizes the previous results that were summarized in Section
\ref{constraints}.

As before, the observables we will use to constrain these models are
the supernova distance moduli as functions of redshift, the CMB power
spectrum on small scales, the local Hubble rate, and the kinematic
Sunyaev-Zel'dovich effect.  The specific data used for each
observable will be explained in the subsections that follow.
We use the parametrized LTB models of
Section \ref{model}, and a Metropolis-Hastings Markov chain Monte
Carlo (MCMC) method to explore parameter space \footnote{Our MCMC code
  is available at \url{http://www.physics.ox.ac.uk/users/bullp}.}.
The likelihood function for each set of parameters is modeled as a
chi-squared distribution, such that $-2 \log \mathcal{L} \approx
\chi^2$, and `goodness of fit' is quantified by comparing to a
$\Lambda$CDM model with $\Omega_\Lambda=0.734$ and $h=0.710$
\cite{WMAPbestfit}.

We first proceed by considering each observable individually, and then
go on to consider the constraints available from combinations of different
observables.  It is found that the additional freedom allowed by
varying the bang time significantly weakens the constraints that each
observable imposes by itself, but that the combined power of all
observables is still enough to effectively rule out these models as a
possible explanation of dark energy.  In particular, we show that neither the
CMB+$H_0$ observations, nor the upper bounds on the kSZ effect for individual 
clusters, have the ability to rule out these models by themselves, as is the case
when the bang time is assumed to be constant.

\subsection{Supernovae}

In the fits that follow we use the Union2 compilation of 557
supernovae, which extends out to $z \sim 1.4$ \cite{Union2}. Other supernova
data sets also exist, and void fitting procedures are known to
exhibit some sensitivity to the data set that is chosen
\cite{CFL,Nadathur}.  We choose the Union2 data as it is the most
extensive catalog, and the most widely used in the literature.
The absolute magnitude of the supernovae in the Union2 data set is an
unknown parameter, and is therefore fitted to each model
individually as a nuisance parameter.  We use the published errors in
this data set, which includes an `intrinsic error' that is added to
minimize the reduced $\chi^2$ of $\Lambda$CDM.  The full Union2
``covariance matrix with systematics'' is used in performing all
of the likelihood estimates that follow.

As with $t_B=$constant, there is no problem fitting the supernova data
without dark energy.

\subsection{The CMB and $H_0$}
\label{results-sn-cmb-h0}

In FLRW cosmology, and for our current purposes, the CMB power spectrum can be 
efficiently specified on small scales with only three pieces of information 
 \footnote{At least five pieces of information are required for a full, 
 relatively model-independent specification of the CMB power spectrum 
 \cite{Vonlanthen}, but in what follows we will choose to marginalize over (or 
 fix) the overall normalization of the power spectrum and the spectral index of 
 the initial scalar power spectrum, as these quantities are not important for 
 constraining the large-scale structure that we are interested in here.}: 
(i) the acoustic horizon scale at decoupling, (ii) the acoustic scale at 
matter-radiation equality, and (iii) the projected scale of the
CMB onto our sky.  This information can be combined into three
parameters in a number of different ways \cite{CR, WangMukherjee, HFZT, Vonlanthen}, but here we choose to specify it as the `shift parameter', $S$, the
Hubble rate at last scattering, $H_*$, and the redshift of the last
scattering surface, $z_*$.  The shift parameter is defined as $S
\equiv d_A(z_*) / \hat{d}_A(z_*)$, where $\hat{d}_A(z_*)$ is the
angular diameter distance to the last scattering surface in a fiducial
spatially flat FLRW model with $\Omega_m \simeq 1$.  This quantity corresponds to the change in
scale of fluctuations on the sky that two observers in different
space-times would see when looking at two identical last
scattering surfaces.

\begin{figure}[tbh]
  \centering
  \includegraphics[scale=0.38]{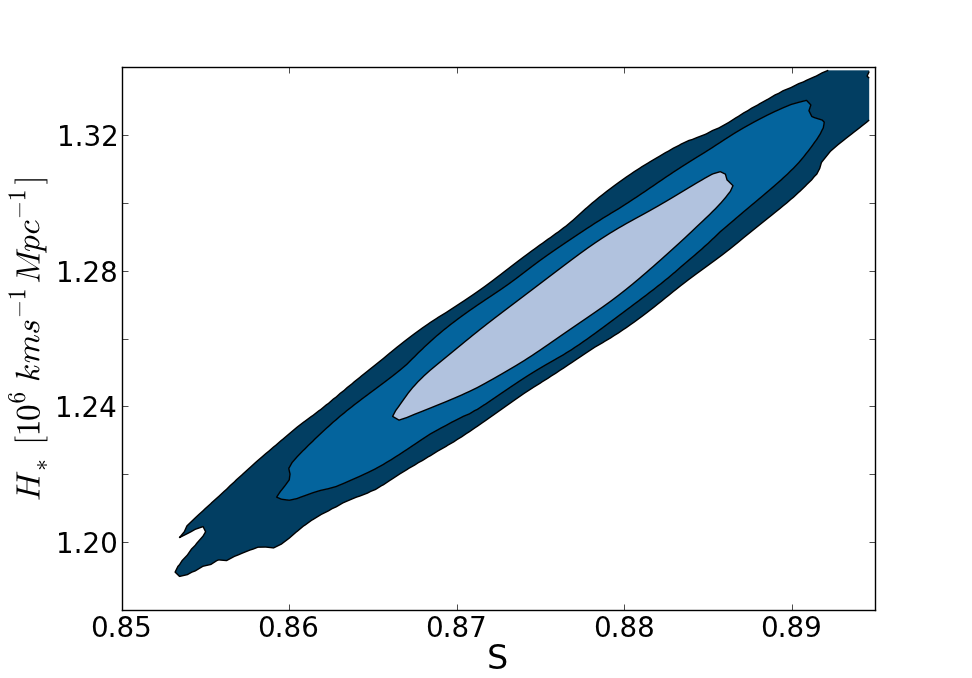}
  \caption{Marginalized likelihood for the parameters $S$ and $H_*$, found using WMAP 7-year
  data and a modified version of CosmoMC \cite{WMAP7}.  Shaded regions show the
  68\%, 95\%, and 99.7\% confidence regions.}
  \label{fig-shift-constraints}
\end{figure}

\begin{figure*}[htb]
\hspace{-50pt}
\includegraphics[scale=0.45]{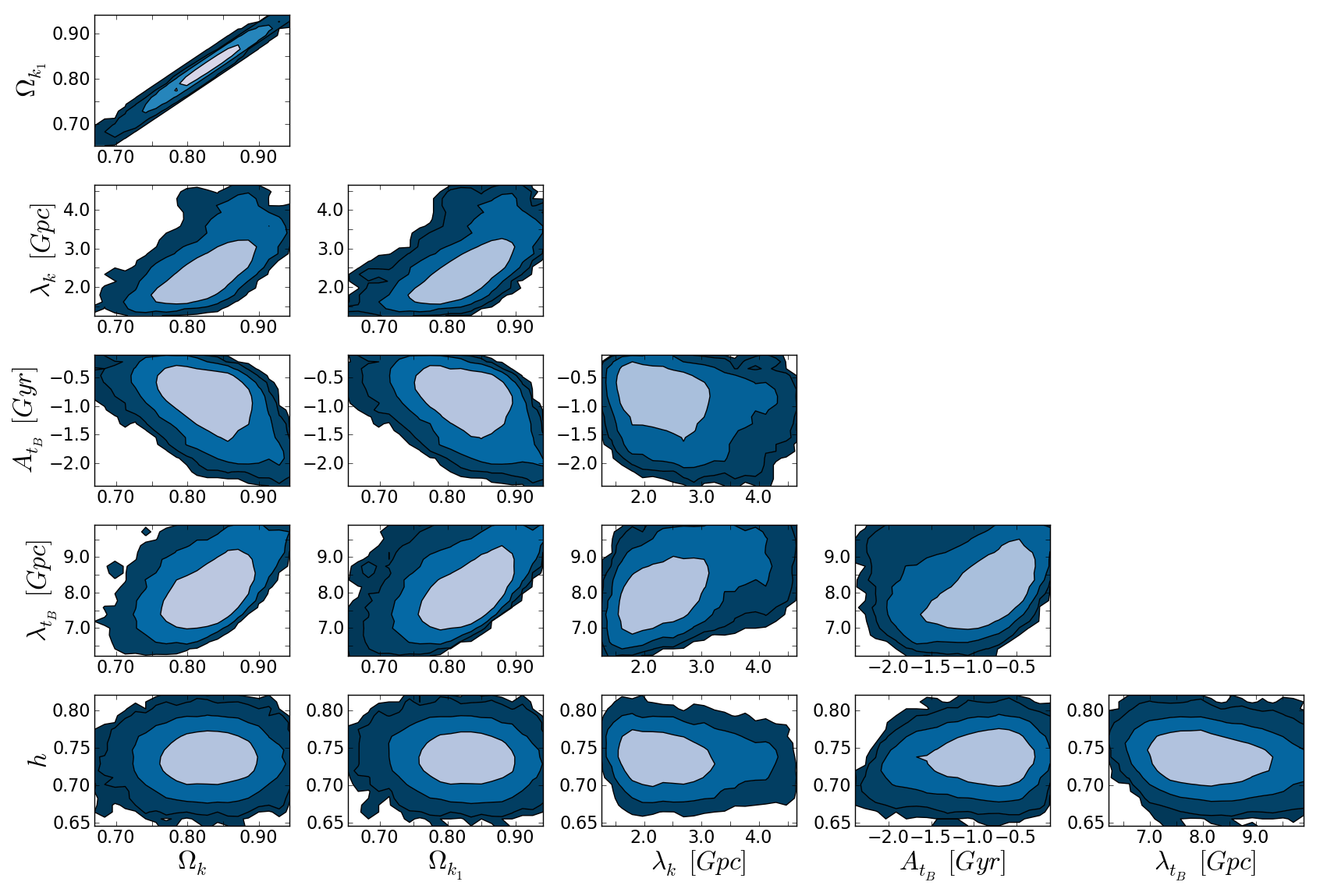}
\caption{Marginalized likelihoods for a Gaussian void model with
   varying bang time.  Shaded regions show the 68\%, 95\%, and 99.7\%
   confidence regions. Here we use $H_0 = 100 h$
   kms$^{-1}$Mpc$^{-1}$. See the text for other definitions.}  
  \label{fig-sn-cmb-h0}
\end{figure*}

Here, for simplicity, we take  $z_* = 1090$, which is the redshift to
the last scattering surface in $\Lambda$CDM and various other models \cite{Vonlanthen} 
\footnote{Allowing $z_*$
  to be free would add an extra nuisance parameter to be fitted for
  in our MCMC runs, which would inevitably loosen the resulting
  constraints. As such, the CMB constraints we derive here should be
  considered to be conservative.}.
It now remains to impose constraints on $S$ and $H_*$.  To do this, we
enforce the condition that the region of space in which the last scattering surface
forms is well-approximated as being homogeneous and isotropic, so that
standard results from FLRW cosmology can be applied in dealing with
all of the physics up until the formation of the last scattering surface.  We can then use
CosmoMC \cite{CosmoMC} to calculate $H_*$ and $d_A(z_*)$ for an
observer in a spatially flat and dust dominated FLRW universe looking
at this surface.  Using the space-time geometry of our void models we can
then calculate $S$ and $H_0$ for an observer at the center of the void
looking at an identical last scattering surface, with identical Hubble rate at last
scattering, and at an identical redshift.  This is the procedure
followed in \cite{CFZ}.

We use the WMAP 7-year data \cite{WMAP7}, with a modified version of
CosmoMC, to constrain our models.  In this analysis we choose to only
use data at $\ell \ge 100$, as the low-$\ell$ power spectrum is
dominated by the ISW effect (see Section
\ref{obs-others} for a brief discussion of this).  This choice weakens
the constraints that can be achieved on the scalar spectral index of
the initial power spectrum, $n_s$.  Conservatively, we fix
$n_s=0.96$ here \footnote{Modifying the form of the initial power spectrum
can significantly weaken CMB constraints on void models \cite{Nadathur}.}.
The constraints that can then be imposed on $S$ and $H_*$ are shown in
Fig. \ref{fig-shift-constraints}.  The best fit values are found to be
$S \simeq 0.875$ and $H_* \simeq 1.27 \times 10^6$ kms$^{-1}$Mpc$^{-1}$.
These values are consistent with those found in \cite{CFZ, ZMS}. Note that 
the CosmoMC CMB fits were not performed jointly with the void model MCMC;
instead, they were run beforehand to get likelihoods for $S$ and $H_*$, 
which we then used as priors for the void model MCMC.

As discussed in Section \ref{constraints-cmb}, it is possible to
construct simple void models with $t_B=$constant that satisfy the
constraints on $S$ displayed in Fig. \ref{fig-shift-constraints}.
This can be achieved by simply changing the spatial curvature of the model
at large $z$ \cite{CFZ}.  The constraints on $H_*$, however, are
more difficult to satisfy. For simple Gaussian voids with $t_B=$constant, 
under the assumptions described above, the WMAP 7-year data \cite{WMAP7}
and the Union2 supernova data set \cite{Union2} are enough to show
that $H_0 \lesssim$ 40 kms$^{-1}$Mpc$^{-1}$ is required, which is in strong
disagreement with the value of $H_0=73.8 \pm 2.4$ kms$^{-1}$Mpc$^{-1}$
found by Riess {\it et al.} \cite{Riess}.  Allowing $z_*$ to vary can
increase the upper bound on $H_0$ by around 5 kms$^{-1}$Mpc$^{-1}$,
and changing the precise functional form of $k(r)$ can also marginally
change $H_0$ (see \cite{Biswas, MZS10}).  These are relatively small
effects, however, and unless one is prepared to reject one or more of
the assumptions given in Section \ref{intro}, models with
$t_B=$constant remain strongly inconsistent with recent measurements
of $H_0$. 

Allowing the bang time function to vary significantly improves the
ability of void models to fit the CMB+$H_0$ data \cite{CFZ}.  In
Fig. \ref{fig-sn-cmb-h0} we show the likelihood plots for the
parameters $(\Omega_k, \Omega_{k_1}, \lambda_k, A_{t_B},
\lambda_{t_B}, H_0)$, when constrained with the WMAP 7-year 
data \cite{WMAP7}, the Union2 data set \cite{Union2}, and the
measurement of $H_0=73.8 \pm 2.4$ kms$^{-1}$Mpc$^{-1}$ \cite{Riess}. 
Good fits to the data are obtained for models with a bang time fluctuation of 
width 8000 Mpc that makes the Universe about 800 million years older in the 
center than it is at large $r$. The curvature profile is narrower than this, 
with a width of 2500 Mpc, and a depth of $\Omega_k=0.83$ at the
center.  The preferred spatial curvature at large radii is 
only $\Omega_{k,2} \sim +0.002$.  It can be seen that in this case the model 
is able to produce an acceptably large value of $H_0$, with a best-fit value of 
73.6 kms$^{-1}$Mpc$^{-1}$.  When compared to the best-fit 
$\Lambda$CDM model we find that $\Lambda$CDM is slightly preferred,
with $\Delta \chi^2 =$+4.5 for 560 degrees of freedom. Most of this difference 
is due to the void model having a poorer fit to the supernova data,
even though it  over-fits $H_0$ and the CMB data.  Voids with slightly more 
complicated spatial curvature profiles produce fits to the data that
are at least as good as $\Lambda$CDM.

In the best fitting models the curvature profile, $k(r)$, is largely
responsible for shaping the void at low redshift, with $z \lesssim 1$. In this
region the bang time gradient is small, and so has little effect.  In
the region $1 \lesssim z \lesssim 2 $ the curvature profile then
flattens out and the bang time gradient begins to change rapidly.
This produces large fluctuations in $H_2(z)$ along our past null cone,
such that $H_2$ can take lower values at large $z$.  The low value of
$H_*$ required at last scattering can then be simultaneously
accommodated with a large value of $H_0$ locally.  The difference in
profile widths therefore helps to explain how it is that a good fit
to the data can be achieved.

\subsection{The kSZ Effect}\label{results-sn-ksz}

Let us now consider the kSZ effect in void models with varying bang
times.  The void-induced dipole, $\Delta T/T$, can be calculated in an
LTB model by following the procedure below:
\begin{enumerate}
 \item On every point on our past null cone, solve the radial null
 geodesic equation for light rays traveling both into and out of the
 center of the void.
 \item Calculate the redshift to the last scattering surface along
 these geodesics using Eq. (\ref{eqn-redshift-central}).
 \item Calculate the dipole, $\Delta T / T$,  using
 Eq. (\ref{ksz-dt}).  This can be converted into an effective
 velocity using $\beta = \Delta T / T$.
\end{enumerate}
This procedure relies on knowing the location of the last scattering
surface at different values of $r$.  For models with a
constant bang time, this surface occurs at a constant time,
$t=t_{LS}$.  In models with varying bang time, however, it will not
occur as a hypersurface of constant $t$, as the presence of a bang
time gradient changes the time evolution of the radial Hubble rate and
density at a given $r$.  We therefore approximate the location of the
last scattering surface as a hypersurface of constant density, $\rho$,
rather than time, $t$.  The precise location of the last scattering
surface will turn out to be important, and we will discuss the
consequences of altering its position as we proceed.

We use the upper limits on $\beta=\Delta T / T$ that
have been measured from nine individual clusters \cite{kSZobs1,
  kSZobs2, kSZobs3}, as collected in \cite{GBH}. These clusters span a
redshift range of $0.18 \leq z \leq 0.55$. The data have
asymmetric statistical errors, and are subject to large systematic
errors of up to $\sim 750$ kms$^{-1}$.  For further
explanation of the uncertainties in this data set the reader is
referred to \cite{GBH}.

Void models with constant $t_B$ have already been shown to be
inconsistent with even this limited data set, as we discussed in
Section \ref{ksztest} \cite{GBH}.  Within this class of models, and subject to
the assumptions outlined in Section \ref{intro}, the best fitting
voids are those that are either very shallow ($\Omega_{k_1} \ll 1$) or
very narrow ($\lambda_k \ll 1$ Gpc), with the latter of these
possibilities only working because it restricts the inhomogeneity to
redshifts at which there are currently no data.
We find that the extra freedom afforded by allowing the bang time to
vary relaxes these tight constraints, and admits the possibility of
allowing $\beta(z)$ to be small even in regions of the Universe that
are strongly inhomogeneous, with $\Omega_{k_1}$ as large as $0.85$.

\begin{figure}[htb]
\hspace*{-25pt}
  \includegraphics[scale=0.42]{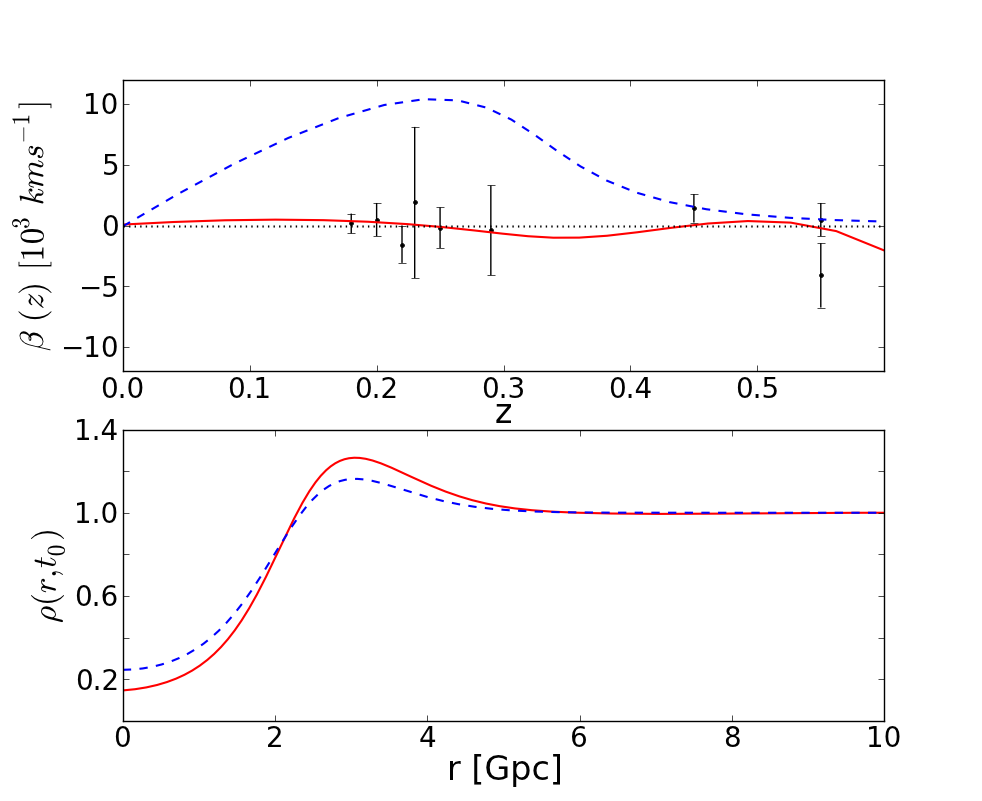}
  \caption{Upper panel: Velocity with respect to the frame where the CMB is 
  isotropic, $\beta(z)$, for models with constant bang time (dashed blue line)
  and non-constant bang time (solid red line).  Both models have the
  same spatial curvature, $k(r)$. The data points are the upper limits
  for the nine clusters used in \cite{GBH}. The solid red curve has a
  bang time fluctuation that is the sum of two Gaussians.  Lower
  panel: Normalized density as a function of $r$ on a hypersurface of constant
  $t$ for the same two models.}
  \label{fig-ksz-comparison}
\end{figure}

Fig. \ref{fig-ksz-comparison} shows an example of a large void with
varying bang time that produces small enough $\beta(z)$ to be
compatible with the data discussed above.  A model with the same
curvature profile, $k(r)$, but a constant bang time is also
displayed.  It can be seen that the additional freedom allowed by
the varying bang time has a considerable impact on $\beta(z)$.  The
energy density profile for this model is also displayed in the figure.
One should note, however, that the functional form of the bang time
fluctuation required to produce this result is more complicated than 
the simple profile of Eq. \ref{eqn-tb-profile}. Instead, a sum of two 
(modified) Gaussian curves was used, of the form
\begin{eqnarray}\label{eqn-ksz-complicated}
t_B(r) = & A_1 \exp \left ( -{\left [(r- r_1)^2/\lambda^2_1 + (r- r_1)^{6}/\lambda^6_1\right ]} \right ) & \\ \nonumber
+ & A_2 \exp \left ( -{\left [(r- r_2)^2/\lambda^2_2 + (r- r_2)^{6}/\lambda^6_2\right ]} \right ), &
\end{eqnarray}
where $A_{1,2}$, $r_{1,2}$, and $\lambda_{1,2}$ are the amplitude, offset 
from the origin and width of the Gaussians respectively. Despite the greatly 
increased freedom in this bang time profile, we were unable to find a model 
consistent with the kSZ and supernova data simultaneously; the best-fit model 
had $\Delta \chi^2_{SN} = 44.1$ with $\Lambda$CDM. It is plausible that the 
situation could be improved by considering yet more complicated functional 
forms for $k(r)$ and $t_B(r)$, a possibility that we investigate in 
Section \ref{results-complicated}. For simplicity, we 
have assumed in Fig. \ref{fig-ksz-comparison} that the offset caused by 
systematic errors in the data is zero.  In reality, the data points would 
likely move slightly toward the curve to which they are being fitted, in
order to improve the likelihood.

Finding models that agree with the upper limits on the statistical kSZ
from ACT and SPT is a more difficult task. The kSZ power spectrum
given by $\Delta T^2$, in Eq. (\ref{eqn-ksz-power}), depends on an
integral of $\beta(z)$ over redshift.  Deviations from $\beta = 0$ at
any redshift therefore accumulate, potentially producing a large kSZ
signal. It is possible that $\beta(z)$ could be made to change sign so
that negative contributions cancel the positive ones, but this would
require a delicate balancing of the competing effects to satisfy the
statistical and single-cluster kSZ data simultaneously. One should bear in
mind, however, that at large enough $z$, the effect of the void on the
observed kSZ effect will decrease, as the angle subtended by the void on the 
distant observer's sky decreases, and the power in the dipole term of the 
anisotropy is shifted to higher multipoles \cite{Alnes}.

In summary, we find that large void models with varying bang times may
have enough extra freedom available to alleviate the constraints that
can currently be imposed from observations of the kSZ effect.  As we discuss in
the following section, however, it is unlikely that after doing this
there will be enough remaining freedom to accommodate any other observables.

\subsection{Combined Constraints \qquad \qquad \qquad (SN+CMB+$H_0$+kSZ)}\label{results-combined}

Let us now consider combining all of the observables we have discussed
so far.  These are the Union2 supernova data, the WMAP 7-year data,
local measurements of $H_0$, and the kSZ effect.

\begin{figure}[htb]
  \hspace*{-10pt}
  \includegraphics[scale=0.45]{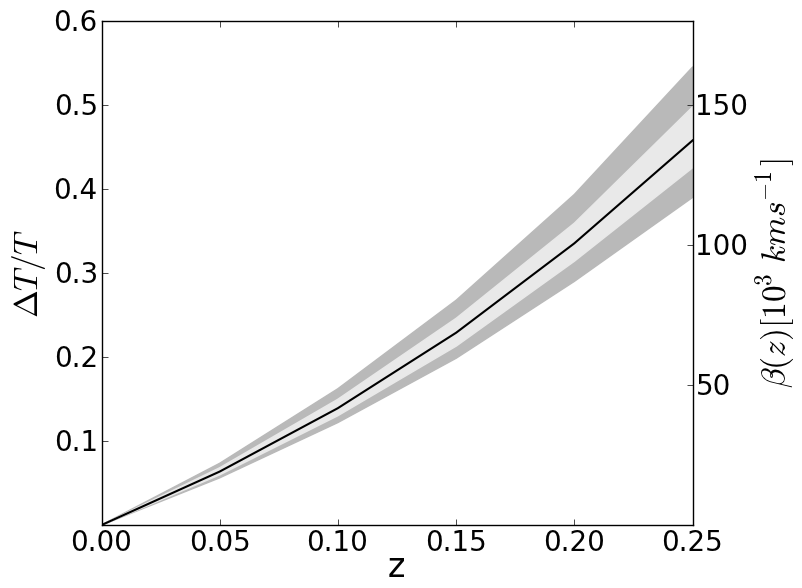}
  \caption{The value of $\Delta T/T$ as a function of redshift for the
  best fitting models to the supernova, CMB and $H_0$ data sets, from
  Section \ref{results-sn-cmb-h0}. The corresponding value of $\beta$ is 
  shown on the right-hand axis. The median and 68\% and 95\% confidence 
  intervals are shown as the black line, and the light gray and dark 
  gray bands, respectively. The actual distribution is bimodal, and here 
  we show only the models with low $\Delta T / T$. Even for low redshifts, 
  $\beta$ is a large fraction of the speed of light.}
  \label{fig-ksz-z}
\end{figure}

\begin{figure}[htb]
  \hspace*{-20pt}
  \includegraphics[scale=0.40]{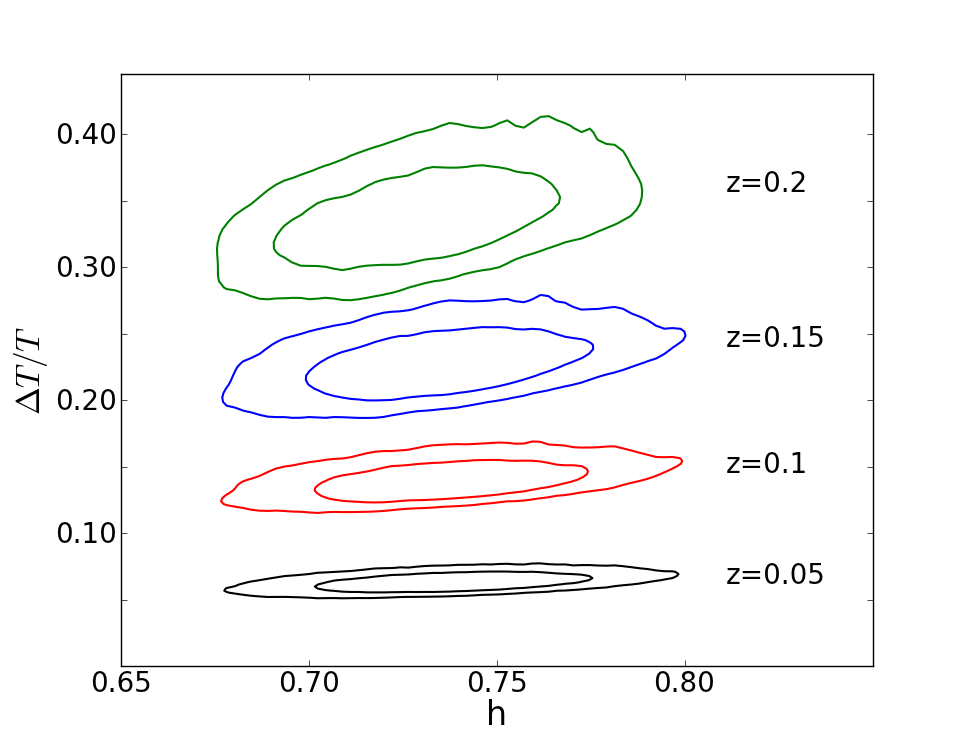}
  \caption{The 68\% and 95\% likelihood contours in the
  space of $\Delta T / T$ and $H_0 = 100 h$ kms$^{-1}$Mpc$^{-1}$, for
  off-center observers at $z=0.05$, $0.10$, $0.15$ and $z=0.20$, for the 
  MCMC sample constrained by SN+CMB+$H_0$ data.} 
  \label{fig-ksz-h0}
\end{figure}

In Fig. \ref{fig-ksz-z} we show the observed value of $\Delta T/T$ as a
function of $z$ that a central observer would measure from the kSZ
effect in the models found in Section \ref{results-sn-cmb-h0}.  These models 
have been shown to provide good fits to the supernova data, and the CMB and
$H_0$ data sets simultaneously.  In Fig. \ref{fig-ksz-h0} we show this
information as a function of $H_0$ for redshifts $z=0.05$, $0.10$, $0.15$, and 
$0.20$. At all redshifts considered the distribution was 
bimodal, with some models having $\beta \approx 1$.  Such incredibly
high velocities are completely inconsistent with the data, and so here
we show only the models with lower $\beta$.
It can be seen that the value of the kSZ signal that
one would observe from the center of these models is extremely large,
even at low redshift.  Such enormous kSZ signals are not compatible
with the data displayed in Fig. \ref{fig-ksz-comparison}, even with
very large additional systematic uncertainties included.

The principal reason for this large effect appears to be the
large width of bang time fluctuation that is favored by the
combination of supernova, CMB and $H_0$ data sets (see
Fig. \ref{fig-sn-cmb-h0}).  Because of this, observers at $z \gtrsim 0.1$
look through regions in which the bang time gradient is large when
they look through the void.  As discussed in Section \ref{tb}, these
regions host shell crossings when $t_B^\prime > 0$.   This
pushes the surface of last scattering to much later times, and causes
significant modifications to the redshift that this surface is seen at
when looking through the center of the void.  The redshift of the last
scattering surface when looking away from the void experiences no such
effect, as it is effectively fixed in position by the CMB data we see
from the center.  As a result, the values of $z_{in}$ and $z_{out}$ in 
Eq. \ref{ksz-dt} differ significantly, and the value of $\Delta T/T$ is 
therefore even larger than in the constant bang time case. Even at low $z$, 
this is unacceptably high.

\begin{figure}[b]
  \hspace*{-15pt}
  \includegraphics[scale=0.33]{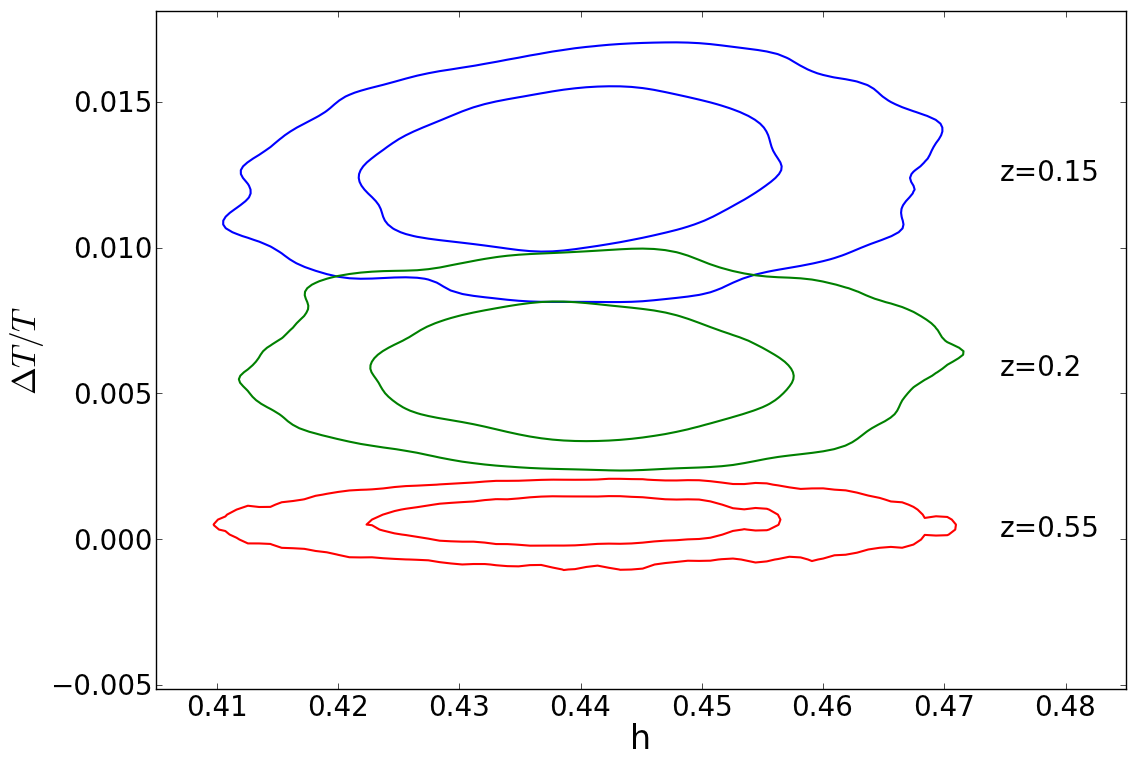}
  \caption{The 68\% and 95\% likelihood contours in the
  space of $\Delta T / T$ and $H_0 = 100 h$ kms$^{-1}$Mpc$^{-1}$, for
  off-center observers at $z=0.15$, $0.20$ and $0.55$, for the MCMC sample 
  constrained by SN+CMB+$H_0$+kSZ data. The $\Delta T / T$ are much lower than 
  for the MCMC sample in Fig. \ref{fig-ksz-h0}, but the Hubble rate is too low 
  to be considered consistent with observations.} 
  \label{fig-ksz-h0-complex}
\end{figure}

\subsection{More Complicated Profiles}\label{results-complicated}

One may now ask whether changing the specific forms of $t_B(r)$ and
$k(r)$ that we have used so far affects our results. We showed in Section 
\ref{results-combined} that very poor agreement with the kSZ data is obtained 
for the models that best-fit the supernova, CMB, and $H_0$ data, but in 
Section \ref{results-sn-ksz} we found that a good fit to the 
kSZ data could be obtained if a more complicated bang time profile was 
used. To see if a good fit to all of the observables is possible with a more 
complicated model, we ran an MCMC simulation using the spatial curvature 
profile of Eq. (\ref{eqn-k-profile}) and the extended bang time profile given by 
Eq. (\ref{eqn-ksz-complicated}). The MCMC was constrained by the supernova, CMB, 
$H_0$, and kSZ data simultaneously.

A plot of kSZ $\Delta T / T$ against $H_0$ for these models is shown in 
Fig. \ref{fig-ksz-h0-complex}, and may be compared with Fig. \ref{fig-ksz-h0}. 
The models that maximize the likelihood have a $\beta(z)$ profile that is 
almost flat over the redshift range of interest (slightly larger at small $z$), 
and much less discrepant with the kSZ data. Relatively narrow spatial curvature 
and bang time profiles are preferred, extending out to only $z \sim 0.2$, and 
the bang time profiles are shifted towards the negative $r$ direction (i.e. 
$r_{1,2} < 0$ in Eq. (\ref{eqn-ksz-complicated})). A preferred $H_0$ of only 
$44.0$ kms$^{-1}$Mpc$^{-1}$ is obtained, and the fit to supernova and CMB data 
is also poor; the best-fit model is inconsistent with the data, with 
$\Delta \chi^2 \approx 270$ compared to $\Lambda$CDM. We conclude that this is 
because the fit is most sensitive to the kSZ data; it is easy to find models 
that are wildly inconsistent with the kSZ data, as evidenced by 
Fig. \ref{fig-ksz-h0}, and so models that minimize the $\chi^2$ with the 
kSZ data above all else are preferred. These tend to have low Hubble rates and 
narrow density profiles, features that are difficult to reconcile with the 
supernova and $H_0$ data. As such, it seems that even with the significantly 
more complex bang time profile, a good fit to all of the data simultaneously 
is not possible.

\begin{figure}[t]
  \hspace*{-27pt}
  \includegraphics[scale=0.42]{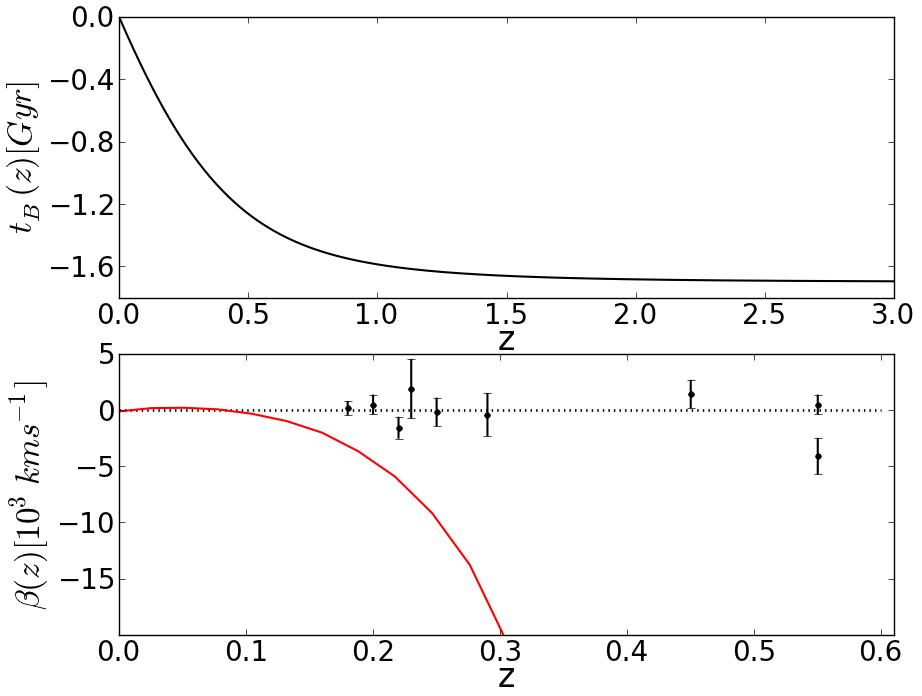}
  \caption{Upper panel: Bang time function, $t_B(r)$, of the LTB model found 
  in \cite{Bolejko2} which reproduces the Hubble rate, $H(z)$, and luminosity 
  distance, $d_L(z)$, of $\Lambda$CDM.
  Lower panel: Velocity with respect to frame in which the CMB is isotropic, 
  $\beta(z)$, for 
  the same LTB model. $|\beta(z)|$ rapidly becomes very large 
  ($\beta\approx-c$ for $z \gtrsim 0.6$), and produces a very poor fit to the 
  kSZ data.}
  \label{fig-cbk-ksz}
\end{figure}

To further investigate the sensitivity of our results to the choice of 
profile parametrization, we now consider the model found by C\'{e}l\'{e}rier
{\it et al.} in \cite{Bolejko2} that was constructed to reproduce the
$\Lambda$CDM values of luminosity distance and Hubble rate as a
function of redshift, but without dark energy.  The density profile
on a hypersurface of constant $t$ takes the form of a ``hump'' in this
model, rather than a void, and the bang time gradient at low $z$ is
negative, so there are no shell crossings at early times.  Instead,
this model has double-valued redshifts at high $z$ (see Section
\ref{tb}), which almost always result in large $\Delta T/T$ because 
$z_{in}$ and $z_{out}$ in Eq. \ref{ksz-dt} differ significantly.  As such, 
this model is also strongly disfavored by current kSZ data (see Fig. 
\ref{fig-cbk-ksz}).

In fact, for large fluctuations in the bang time function we expect that there 
will \textit{always} be a large dipole seen by off-center observers at some 
range of redshifts, corresponding to lines of sight that pass near to regions 
with a non-zero bang time gradient
  \footnote{While the model in Fig. \ref{fig-ksz-comparison} can be seen to fit 
  current kSZ data, these data only extend out to $z \sim 0.6$. Even this model 
  will have a large $\beta(z)$ for some $z > 0.6$.}.
It therefore appears that one cannot simultaneously fit the supernova,
CMB, $H_0$ and kSZ observations with a single LTB model unless one is
prepared to violate one or more of the assumptions made in Section \ref{intro}.

\section{Discussion}\label{discussion}

In order to rigorously establish that $\Lambda \neq 0$, and that the
concordance model of cosmology is correct, a model is required within
which observations can be interpreted. The homogeneity and isotropy of
the Universe on large scales is often assumed, and most analyses are
performed using the highly symmetric FLRW solutions of general
relativity. These are not the only viable ways to model the Universe,
however, and recent advances in observational cosmology allow us to
empirically test alternatives rather than relying on assumed
symmetries of space-time on the largest scales. In this paper we used
the spherically-symmetric, dust-only Lema\^itre-Tolman-Bondi (LTB)
class of general relativistic cosmological models, in their full
generality, to test the radial homogeneity of the Universe on large
scales. In particular, we consider the magnitude of the kinematic
Sunyaev-Zel'dovich (kSZ) effect, which measures the dipole anisotropy
of the CMB through a shift in the spectrum of CMB photons reflected
from hot gas in clusters of galaxies. The kSZ effect is sensitive to
large-scale inhomogeneity, as observers inhabiting an inhomogeneous
universe would generically expect to see large anisotropies on their
CMB skies.

In Section \ref{model}, we introduced the theoretical framework for
specifying these models and calculating observables in them. We
defined a simple parametrization of the LTB radial function $k(r)$
that governs radial inhomogeneity at late times, and the function
$t_B(r)$ that governs it at early times. This has been used to
investigate underdense `voids' which have previously been shown to
produce good fits to the supernova data. We have discussed potential
problems with models that are inhomogeneous at early times (i.e. that
have non-constant $t_B$), including the potential for disruptions
around our observed surface of last scattering. 
In models with a positive radial derivative of the bang
time, $t_B^\prime >0$, it is found that shell crossing singularities
form at early times, which pushes the surface of last scattering to
later times. In models with negative bang time gradient, $t_B^\prime <
0$, regions with a negative radial Hubble rate, $H_2 <0$, form and the
distance-redshift relation, $r(z)$, ceases to be monotonic. These
features have observable effects that ultimately lead to predictions
of a large CMB dipole anisotropy at low redshifts.

In Section \ref{constraints} we reviewed some of the observational
constraints that can be imposed on void models with constant bang
time. Three key sets of observables were considered: The  distance
moduli of supernovae, the small-angle CMB power spectrum plus local
Hubble rate, and upper limits on the magnitude of the kSZ effect for
individual clusters of galaxies. Voids can fit the supernova data
easily, but are unable to fit recent measurements of the CMB and $H_0$
simultaneously (they predict a value of $H_0$ that is far too
low). Similarly, voids which fit the supernova data predict a large
CMB dipole at redshifts up to $z \sim 1$, which is inconsistent with
current kSZ measurements.

In Section \ref{results} we considered the effect on these constraints
of allowing the bang time to vary. This resulted in a significant
increase in the freedom of the models, and allowed the supernovae and
CMB+$H_0$ data sets to be fit simultaneously, even with our simple
parametrization of the LTB radial functions. Models with small kSZ
signals, consistent with the data, were also found, but these required
more complex profiles and gave worse fits to the supernova data.  We
then proceeded to combine all of the observational constraints, and
found that voids which are able to fit the supernovae, CMB and $H_0$
predict an extremely large kSZ effect which is orders of magnitude
greater than the measured upper limits. A joint fit to the supernova, 
CMB, $H_0$, and kSZ data with a significantly more complicated bang 
time profile also failed to produce good agreement with the data.

We also argued that any void model with a significant bang time inhomogeneity 
will produce a large kSZ effect at {\it some} redshift. Given that a varying 
bang time is necessary to resolve the low-$H_0$ problem, it seems that the
combination of supernovae, CMB+$H_0$ and kSZ data is enough to
effectively rule out LTB void models that attempt to explain
cosmological data without dark energy, subject to the assumptions made
in Section \ref{intro}.  This goes some way toward demonstrating the
homogeneity of the Universe on large scales.

We used the dipole approximation to calculate the magnitude of the kSZ effect 
in our models, but this only holds if the dipole term dominates the anisotropy 
of an off-center observer's sky \cite{Alnes, GBH}. Otherwise, higher multipoles 
become important, and the dipole approximation overestimates the kSZ effect. 
The dipole will dominate as long as most lines of sight on the observer's sky 
pass through the void, as will be the case if the observer is firmly inside 
it, for example. For the models in Section \ref{results-sn-ksz} and 
\ref{results-combined}, and the C\'{e}l\'{e}rier {\it et al.} model in 
Section \ref{results-complicated}, the inhomogeneity extends out to $z \ge 2$, 
and so the dipole approximation will always be a good one for observers at 
$z < 0.6$, where the kSZ data lie. The models with more complicated bang time 
profiles considered in Section \ref{results-complicated} have much narrower inhomogeneities, however, and so we would expect the dipole approximation to 
be worse. These models predict low kSZ magnitudes and are close to being 
consistent with the kSZ data, so any overestimate due to the dipole 
approximation will have little effect on their total $\chi^2$, which is anyway 
dominated by the poor fits to the supernova, CMB and $H_0$ data. We therefore 
conclude that the dipole approximation is sufficient for our purposes.

As we have tried to make clear throughout, our results are subject to
several caveats that are summarized in Section \ref{intro}. The
first, that we are exactly in the center of a perfectly spherically
symmetric void, serves to simplify our calculations but is clearly
unrealistic as it fails to take into account angular variations in,
for example, the galaxy distribution. This could affect observables
such as the dipole anisotropy of the CMB seen by off-center observers,
potentially weakening the constraints we have derived using the kSZ
effect. Considering ourselves as off-center observers \cite{Alnes,
  Foreman} and introducing linear perturbations \cite{CCF, Zibin}
would produce more realistic models and allow more observational data
to be used (e.g. the matter power spectrum) at the expense of a
significant increase in complexity. A better understanding of linear
perturbations would also go some way toward addressing our second caveat, 
that the formation of the last scattering surface must be in an 
approximately-FLRW region. In general, one could expect features such as 
the coupling of scalar and tensor modes in LTB perturbations to produce 
secondary effects such as large B-mode polarizations \cite{CCF}.  This type 
of effect is completely absent in linear perturbation theory about FLRW
backgrounds.

A particular limitation of LTB solutions as cosmological models is
that they contain only dust, and cease to be applicable when radiation
becomes important. If we want to approximate the Universe as an LTB
model at late times, we must therefore match it to an appropriate
solution containing radiation at early times. 
Solutions involving separate inhomogeneous matter and radiation fluids
\cite{CR}, spatially-varying physical quantities such as the
photon-baryon ratio \cite{YNS}, and scale-dependent initial power
spectra \cite{Nadathur} have been considered, and serve to give some
idea of the extra freedom that might be obtained in more general
models. Specifically, altering the location and properties of the
surface of last scattering can have a profound effect on the predicted
kSZ signal \cite{YNS} and the observed CMB \cite{Nadathur, CR}, and
if one is prepared to consider this additional freedom, then our present 
results should not be expected to hold.  

Finally, let us consider LTB models in the context of general
inhomogeneity. Rather than allowing space-time to be described by a
single LTB metric, it has been suggested that the LTB geometry could
be used as an effective geometry to model the scale dependence of
inhomogeneity after some averaging procedure has been applied to the
fine-grained structure of the actual inhomogeneous geometry of the
real Universe \cite{Celerier11}.  This is a considerable departure
from the situation we have been considering here.  In particular, if other
observers are able to construct similar effective spherically
symmetric geometries about their own locations then we should no
longer expect distant clusters to see a large dipole in their CMB
sky.  This would completely relax the constraints that can be imposed
from observations of the kSZ effect.

\section*{Acknowledgements}

We are grateful to K. Bolejko, C. Clarkson, A. Coley, J. Dunkley, J.P. Zibin, and 
J. Zuntz for helpful comments and discussions, and to the BIPAC, the STFC, and 
the Oxford Martin School for support.
PB and TC would like to thank the Cosmology and Gravity group at the
University of Cape Town, and the General Relativity and Cosmology group
at Dalhousie University for hospitality while some of this work was
performed. \\ 

{\bf Note added:} A preprint by J.P. Zibin \cite{Zibin11} discussing the effect of bang time 
fluctuations on another tracer of anisotropy in voids, Compton $y$-distortion, 
appeared shortly after the original version of this paper was released. Its 
conclusions are in broad agreement with those presented here: Fluctuations 
in the bang time that are large enough to have a significant effect on the 
geometry of the Universe at late times (and thus have any bearing on the 
low-$H_0$ problem) would result in Compton $y$-distortions many times larger 
that can be reconciled with current observational constraints.

\end{document}